\documentclass[onecolumn]{aastex61}

\usepackage{array}
\newcolumntype{Z}[1]{>{\raggedright\let\newline\\\arraybackslash\hspace{0pt}}m{#1}}

\newcommand{\mydeg}{{$^{\circ}$}}

\newcommand{\tgtes}{Lele\={a}k\={u}honua}
\newcommand{\gaia}{{\it Gaia}}

\tosubmitjournal{Astronomical Journal}
\shortauthors{Buie et al.}
\shorttitle{Distant TNO occultation}

\begin{document}

\title{A single-chord stellar occultation by the extreme TNO (541132) \tgtes}

\correspondingauthor{Marc Buie}
\email{buie@boulder.swri.edu}

\author[0000-0003-0854-745X]{Marc W. Buie}
\affil{Southwest Research Institute, 1050 Walnut St., Suite 300, Boulder, CO~~80302}
%\email{buie@boulder.swri.edu}

\author[0000-0002-6477-1360]{Rodrigo Leiva}
\affil{Southwest Research Institute, 1050 Walnut St., Suite 300, Boulder, CO~~80302}
%\email{rleiva@colorado.edu}

\author[0000-0002-0915-4861]{John M. Keller}
\affil{University of Colorado Boulder, 2000 Colorado Ave, Boulder, CO 80309}
%\email{john.m.keller@colorado.edu}

\author[0000-0002-2193-8204]{Josselin Desmars}
\affil{LESIA, Observatoire de Paris, PSL Research University, CNRS,
Sorbonne Universit\'e, UPMC Univ. Paris 06, Univ. Paris Diderot,
Sorbonne Paris Cit\'e}
%\email{josselin.desmars@obspm.fr}

\author[0000-0003-1995-0842]{Bruno Sicardy}
\affil{LESIA, Observatoire de Paris, PSL Research University, CNRS,
Sorbonne Universit\'e, UPMC Univ. Paris 06, Univ. Paris Diderot,
Sorbonne Paris Cit\'e}
%\email{Bruno.Sicardy@obspm.fr}

\author[0000-0001-7032-5255]{JJ Kavelaars}
\affil{National Research Council of Canada, Victoria BC V9E 2E7, Canada}
%email{JJ.Kavelaars@nrc-cnrc.gc.ca }

\author{Terry Bridges}
\affiliation{Dept of Physics and Astronomy,
Okanagan College, Kelowna, BC, Canada}

\author{Robert Weryk}
\affil{University of Hawaii, Manoa}
%\email{weryk@hawaii.edu}

\author{Dave Herald}
\affil{Occultation Section of the Royal Astronomical Society of New Zealand (RASNZ), P.O. Box 3181, Wellington, New Zealand}
\affil{International Occultation Timing Association (IOTA), P.O. Box 7152, Kent, WA 98042, USA}
\affil{Canberra Astronomical Society, Canberra, ACT, Australia}

\author{Sean L. Haley}	
\affil{University of Colorado Boulder, 2000 Colorado Ave, Boulder, CO 80309}

\author{Ryder Strauss}	
\affil{University of Colorado Boulder, 2000 Colorado Ave, Boulder, CO 80309}

\author{Elizabeth Wilde}	
\affil{University of Colorado Boulder, 2000 Colorado Ave, Boulder, CO 80309}

\author{Robert Baker}	
\affil{RECON, Research and Education Collaborative Occultation Network, USA}	
\affil{Wildwood Institute for STEM Research and Development, Los Angeles, California, USA}

\author{Ken Conway}	
\affil{RECON, Research and Education Collaborative Occultation Network, USA}	
\affil{Foothills Library, Yuma, Arizona, USA}

\author{Bryan Dean}	
\affil{RECON, Research and Education Collaborative Occultation Network, USA}	

\author{Mackenzie Dunham}	
\affil{CanCON, Canadian Research and Education Collaborative Occultation Network, Canada}
\affil{Penticton Secondary School, Penticton, British Columbia, Canada}

\author{James J Estes}	
\affil{RECON, Research and Education Collaborative Occultation Network, USA}	
\affil{Laughlin HS/Bullhead City, Laughlin, Nevada, USA}

\author{Naemi Fiechter}	
\affil{RECON, Research and Education Collaborative Occultation Network, USA}	
\affil{Penticton Secondary School, Penticton, British Columbia, Canada}

\author{Rima Givot}	
\affil{RECON, Research and Education Collaborative Occultation Network, USA}	
\affil{Sisters High School, Sisters, Oregon, USA}

\author{Cameron Glibbery}	
\affil{CanCON, Canadian Research and Education Collaborative Occultation Network, Canada}
\affil{Penticton Secondary School, Penticton, British Columbia, Canada}

\author{Bruce Gowe}	
\affil{CanCON, Canadian Research and Education Collaborative Occultation Network, Canada}	
\affil{Penticton Secondary School, Penticton, British Columbia, Canada}

\author{Jennifer N. Hayman}	
\affil{CanCON, Canadian Research and Education Collaborative Occultation Network, Canada}
\affil{Penticton Secondary School, Penticton, British Columbia, Canada}

\author{Olivia L Ireland }	
\affil{CanCON, Canadian Research and Education Collaborative Occultation Network, Canada}
\affil{Penticton Secondary School, Penticton, British Columbia, Canada}

\author{Matthew Kehrli}	
\affil{RECON, Research and Education Collaborative Occultation Network, USA}	
\affil{California Polytechnic State University, San Luis Obispo, California, USA}

\author{Erik M. Moore }	
\affil{CanCON, Canadian Research and Education Collaborative Occultation Network, Canada}
\affil{Penticton Secondary School, Penticton, British Columbia, Canada}

\author{Matthew A. MacDonald}	
\affil{CanCON, Canadian Research and Education Collaborative Occultation Network, Canada}
\affil{Penticton Secondary School, Penticton, British Columbia, Canada}

\author{Delsie McCrystal}	
\affil{RECON, Research and Education Collaborative Occultation Network, USA}	
\affil{Sisters High School, Sisters, Oregon, USA}

\author{Paola Mendoza}	
\affil{RECON, Research and Education Collaborative Occultation Network, USA}	

\author{Bruce Palmquist}	
\affil{RECON, Research and Education Collaborative Occultation Network, USA}	
\affil{Central Washington University, Ellensburg, WA, USA}

\author{Sherry Rennau}	
\affil{RECON, Research and Education Collaborative Occultation Network, USA}	
\affil{Parker High School, Parker, AZ, USA}

\author{Ramsey Schar}	
\affil{RECON, Research and Education Collaborative Occultation Network, USA}	
\affil{Sisters High School, Sisters, Oregon, USA}

\author{Diana J. Swanson}	
\affil{RECON, Research and Education Collaborative Occultation Network, USA}	
\affil{California Polytechnic State University, San Luis Obispo, California, USA}

\author{Emma D. Terris}	
\affil{CanCON, Canadian Research and Education Collaborative Occultation Network, Canada}
\affil{Penticton Secondary School, Penticton, British Columbia, Canada}

\author{Holly Werts}	
\affil{RECON, Research and Education Collaborative Occultation Network, USA}

\author{J. A. Wise}	
\affil{RECON, Research and Education Collaborative Occultation Network, USA}	
\affil{Wildwood Institute for STEM Research and Development, Los Angeles, California, USA}

\begin{abstract}

A stellar occultation by the extreme, large perihelion, trans-Neptunian
object (541132) \tgtes\ (also known by the provisional designation of 2015~TG$_{387}$)
was predicted by the Lucky Star project and observed with the Research and Education
Collaborative Occultation Network (RECON) on 2018 October 20 UT.  A single detection
and a nearby non-detection provide constraints for the size and albedo.  Assuming
a circular profile, the radius is $r=110^{+14}_{-10}$~km corresponding to
a geometric albedo $p_V=0.21^{+0.03}_{-0.05}$, for an adopted absolute magnitude
of $H_V=5.6$, typical of other objects in dynamically similar orbits.  The
occultation also provides a high-precision astrometric constraint.

\end{abstract}

\section{Introduction}

The recently discovered trans-Neptunian object (TNO), \tgtes, is a dynamically
interesting object that is in one of the
more extreme outer solar system orbits known so far \citep{Sheppard2019}.
\tgtes\ has an absolute magnitude of $H_V \sim 5.6$ that puts it in the top
10\% of all TNOs and Centaurs and in the top 20\% of all scattered disk objects
in absolute magnitude.
Its osculating orbit at the occultation epoch had a semi-major axis of
1019~AU and an eccentricity of 0.936 giving it a perihelion distance of 65~AU,
aphelion distance of 1972~AU, orbital inclination of 11.7\mydeg, and orbital
period of more than 32,500 years.
The most interesting aspect of the orbit is its perihelion distance, putting it
well beyond all of the known perturbers in the solar system.  This object
is similar in orbital properties to Sedna, another distant large
perihelion distance object ($a=479$~AU, $e=0.841$, $i=11.9$\mydeg, $q=76$~AU).

\tgtes\ is presently faint but still within reach of large telescopes at
a magnitude of $V=24.6$ at a heliocentric distance of 78.5~AU.  At
aphelion, this object would be all but undetectable at magnitude $V=38$.  Little
is known about this object \citep{Sheppard2019} but the absolute magnitude
provides a size constraint.  A 4\% albedo would imply a diameter of 510~km
while a 30\% albedo would imply a diameter of 180~km, where the range is motivated by the albedo estimates of known TNOs \citep{Kovalenko2017a}.  Despite only having
a 3-year astrometric observational arc at the time of occultation, its positional error was low enough to
make it viable target for study via stellar occultation.

\section{Observations}

The observations for this occultation campaign were part of the RECON
project \citep{Buie2016} in collaboration with the Lucky Star project
\citep[cf.,][]{Berard2017, Leiva2017, Ortiz2017}.  At the time of the
event, the cross-track uncertainty was 8000~km (0.14 arcsec), the 1-$\sigma$ timing
uncertainty was 5~minutes, and the probability of success was estimated
to be 9\%
based on the RECON prediction and participation by {\em all} teams.
The network
covered from $-0.11\sigma$ to $0.09\sigma$ in the cross-track direction.
In the downtrack direction we planned to cover $\pm2.5\sigma$ (30 minutes).
Normal operations for RECON require a 30-day notice for a full
campaign but the notification for this event was too late.  In this case,
we put out an optional call -- meaning teams were encouraged to observe
at their discretion.  With a cross-track uncertainty of $\pm$8000~km, no particular
spot on Earth was preferred even though our network straddled the predicted
centerline.  Any site able to see the target star was a useful observing station.

\begin{center}
\includegraphics[scale=3.0]{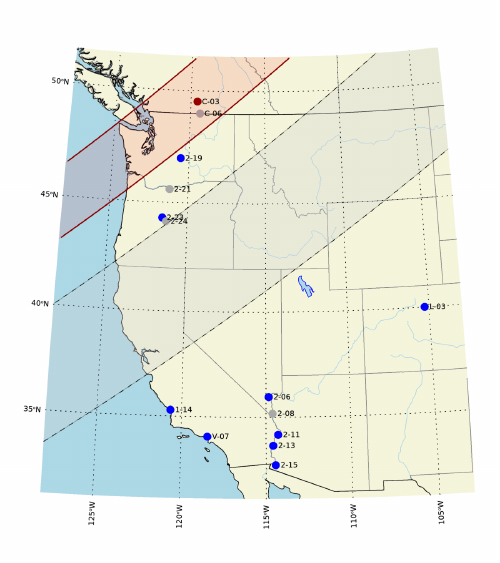}
\figcaption{\label{fig-map}
Map showing locations of observing stations and ground-tracks.
The labeled symbols indicate the positions of the observations
(see Table~\ref{tbl-sites} and Fig.~\ref{fig-lc} for more information about
the sites and their data).  The pair of dashed lines with the gray
shaded region shows the RECON
prediction using a diameter of 480~km (5\% albedo).  The pair of red
lines with red shading indicates the actual track and the derived diameter of 220~km
(see \S\ref{sec:results}).  The prediction uncertainty was much too large
to be shown on this map.
}
\end{center}

The Lucky Star prediction for the event had a geocentric mid-time of
2018-10-20 05:08:02 UTC. The RECON prediction had a later mid-time of
2018-10-20 05:22:31 UTC, a difference of almost 3$\sigma$.  This difference
arises from the astrometry weighting scheme used by Lucky Star
\citep{Desmars2015} compared to RECON where they give the same weight to all astrometric
data \citep{Buie2016}. We chose to base our observations plan on the RECON
prediction because time was short and all the planning information was
already available in a form usable by our teams.  The RECON prediction (dashed
line with grey shading)
and the participating sites are shown in Fig.~\ref{fig-map}.  The color coding
of the event sites uses dark red to
indicate a positive detection, blue for data that show no events,
and grey for sites that tried but were unable to collect constraining
data either positive or negative (additional details are provided in Table~\ref{tbl-sites}).

At the time of the event, the object was 78.7 AU from the Sun
and 77.7 AU from the Earth.  The TNO was moving 24.0 km/sec on the sky
relative to the star (1.5 arcsec/hr).  The sky-plane image scale was
56,000 km/arcsec.  The \gaia\ DR2 \citep{gaia2018} position of the star at
the epoch of the appulse corrected for parallax was RA=00:12:17.891651, Dec=$+$13:15:07.49674, J2000 (\gaia\ DR2 source ID 2767694522024100608).
The uncertainty in the star position, dominated by the proper motion
uncertainty, was (0.33, 0.31) mas. These uncertainties include the estimated systematic uncertainties for the parallax and proper motion from \cite{Lindegren2018} which we add in quadrature. The star has a catalog magnitude of $G=14.5$.
For this event, essentially all of the ground-track uncertainty was from the
uncertainty in the orbit of the object.
At the time of observation, the 81\% illuminated Moon was just 36\mydeg\
away from the target star.  The Sun was well below the horizon for all
teams, ranging from 41\mydeg\ to 54\mydeg\ below the local horizon across
the network.  The target altitude was high enough, ranging from 52\mydeg\ to
69\mydeg\ above the horizon, that atmospheric attenuation was not a concern.

Fourteen of the RECON stations attempted the event and of these, ten collected
constraining data.  From the other four sites, one had telescope alignment
issues, two were not able to point to the correct field, and one recorded data
only during a portion of the planned time window and started too late.  The weather was bad in
the middle of the network, preventing any observations there.
The observational details are
summarized in Table~\ref{tbl-sites}.  Each entry in the table indicates the
team ID, summarized in \citet{Buie2016} and cross-referenced on Fig.~\ref{fig-lc}.
Note that the ``C'' codes are for new teams recently added to the network
that are sited in southern British Columbia, Canada.  The start and ending
times of the data recordings are listed.  If no times are listed that team
was unable to collect data.  The column labeled ``SUP'' gives the camera
``sense-up'' setting, roughly equal to an exposure time of SUP/60 seconds,
see \citet{Buie2016} for details on sense-up.  The entries listed for the
Canadian sites are actual exposure times.  The Canadian sites do not
use the standard detector-telescope setup described in \citet{Buie2016} and
can directly set a specific exposure time.  The observing locations are then
given with the latitude and longitude (in degrees) and the altitude (in meters)
on the WGS84 datum.  The ``Q'' column is an indication of the data quality.
Those without data recorded but workable sky conditions are shown with 0.
The range of 1-5 gives the quality ranging from 1 (lowest) to 5 (best).
The observers involved are listed followed by relevant comments regarding the
data.

The Canadian sites use a different camera from the other RECON sites
(described in \citet{Buie2016}).  The new camera is a QHY174M-GPS (hereafter
referred to as QHY).  This
device is based on an sCMOS detector with 1920 by 1200 pixels and a 1~ms
readout time and has no mechanical shutter.
The images from the QHY cameras were stored directly as FITS image files
unlike the video capture data from the usual RECON systems.  More importantly,
this new camera has an integrated GPS receiver and the starting times of
each image, good to better than 1~ms, are stored with each frame. Penticton
(C-03) used a Meade LX200 30~cm f/10 reduced to f/6.3 giving an effective
scale of 0.7\arcsec/pixel (camera binned 2x2) and a field of view of
11.6\arcmin\ x 7.2\arcmin. Image capture was made with SharpCap software
Version 3.1.5214 using a software gain=360 giving an effective gain of
0.116~e$^-$/ADU and read-out noise of 2.5~e$^-$. Anarchist Mt. Observatory (C-06)
site uses a 30~cm f/4.9 Cassegrain astrograph telescope giving a scale
of 0.8\arcsec/pixel and a field of view of 26.3\arcmin\ x 16.5\arcmin.
Image capture was made with SharpCap software Version 3.1.5220.0
using a software gain=400 giving an effective gain of 0.067~e$^-$/ADU
and read-out noise of 2.3~e$^-$.

\begin{longrotatetable}
\begin{deluxetable}{lcccccccp{5cm}p{5cm}}
\tablecaption{Participating Sites\label{tbl-sites}}
\tablewidth{0pt}
\tablehead{
\colhead{SiteID}&
\colhead{UT start}&
\colhead{UT end}&
\colhead{SUP}&
\colhead{Lat}&
\colhead{Lon}&
\colhead{Alt}&
\colhead{Q}&
\colhead{Observers}&
\colhead{Comment}
}
\startdata
1-14 CPSLO&05:07:35&05:37:44&128& +35.300508& $-$120.659893& 84& 3& D. Swanson, M. Kehrli& \\
2-06 Searchlight&05:07:05&05:37:00&128& +35.965350& $-$114.836775& 695& 5& C. Wiesenborn& \\
2-08 Laughlin&05:07:15&05:37:13&128& +35.162658& $-$114.610642& 251& 1& J. Estes, M. Cordero, D.L. Estes, M. Cruz, A, Magaw& Wrong field recorded.\\
2-11 Parker&05:07:03&05:37:08&128& +34.141088& $-$114.288362& 98& 5& S. Rennau, R. Reaves& \\
2-13 Blythe&05:07:18&05:37:20&128& +33.607967& $-$114.577870& 51& 5& D. Barrows, N. Patel& \\
2-15 Yuma&05:07:11&05:37:20&128& +32.659458& $-$114.436203& 65& 4& K. Conway, D. Conway& Vibrations in first 4 minutes.\\
2-19 Ellensburg&05:05:10&05:30:04&128& +47.002200& $-$120.540112& 489& 3& B. Palmquist, R. Palmquist& Vibrations during the capture.\\
2-21 The Dalles&\nodata&\nodata&\nodata& (+45.596173)& ($-$121.188597)& (77)& 0& B. Dean, M. Dean& Technical issues.\\
2-23 Sisters&05:07:02&05:37:15&128& +44.296303& $-$121.577402& 968& 4& D. McCrystal, R. Schar, R. Givot, P. Mendoza, H. Werts, K. Werts& \\
2-24 Bend&05:07:04&05:37:42&128& +44.132702& $-$121.331668& 974& 1& L. Matheny. A.-M. Eklund, R. Crawford.& Wrong field recorded.\\
C-03 Penticton&05:05:00&05:37:55&4& +49.533883& $-$119.557500& 470& 5& B. Gowe, M. Dunham, J. Hayman, M. MacDonald, E. Moore, N. Fiechter& \\
C-06 Anarchist&05:15:45&05:33:11&0.5& +49.008827& $-$119.362968& 1087& 5& P. Ceravolo, D. Ceravolo& Recording started late.\\
L-03 SwRI&05:06:14&05:35:25&128& +40.003602& $-$105.262798& 1642& 3& J. Keller, R. Leiva, L. Wilde, R. Strauss, S. Haley& \\
V-07 Wildwood&05:07:01&05:40:31&90& +34.033833& $-$118.451282& 24& 4& I. Turk, T.-D. Brown, I. Norfolf, R. Baker, J. Wise& Camera working abnormally.\\
\enddata
\tablecomments{\scriptsize
All site locations are referenced to the WGS84 datum.  Positions for
sites with no data report the nominal team location (shown with enclosed
parentheses) and the team leader(s). The entries in SUP column for C-03 and C-06 are
actual exposure times in seconds.
}
\end{deluxetable}

\end{longrotatetable}
%Table requires some edits from the computer generated version
%  C-06, exposure time to 0.5
% shorten names (laughlin, boulder city, Anarchist)
% change - to $-$

Despite the sparse coverage from the participating sites and the large
uncertainty, a positive occultation detection was recorded at Penticton
(C-03) with an additional nearby miss at Ellensburg (2-19).  Given the
site locations involved, the probability of getting one or more chords
was 6\% and there was an equal chance of getting
just one chord compared to getting multiple chords.

\section{Data Analysis}

The extraction of timing and lightcurve data depends somewhat on the type of
system used.  Details regarding the RECON video systems can be found in
\citet{Buie2016} and \citet{Benedetti-Rossi2016}.  In the latter reference,
the most relevant information is in \S{3.1.2}.  All of the datasets other
than Penticton (C-03) and Anarchist Mountain Observatory (C-06) were taken
with the standard RECON video-based system.  The Wildwood (V-07) system is
very similar to the standard but was not provided as a package of gear
from RECON.  Instead, the Wildwood team procured their own equipment when
they joined the project and it is not precisely the same as the original.
As a result, the Wildwood data required additional analysis due to a higher
level of variable background video noise.
The resulting lightcurves are shown in Fig.~\ref{fig-lc} and the details of the
data reduction process is given below.

\subsection{Video data}

The video processing starts with extraction of time from the IOTA-VTI
timestamps superimposed on each video field.  Second, the video field
order is determined so that the correct video frames are assembled.
The integration boundaries are then identified and the sets of frames are
averaged to a single image per integration with an integration mid-time
determined from the timestamps
\citep[see details in][]{Buie2016,Benedetti-Rossi2016}.  The images have
a dark/bias subtracted in this process.  At this point, the series of
images is processed with standard image analysis methods.

\subsection{sCMOS imager data}

The Penticton (C-03) and Anarchist Mt.\null\ Observatory (C-06) teams used a new
model of sCMOS camera, the QHY174M-GPS that was first used for the occultation
results on (486958) Arrokoth \citep{Buie2020} and from that experience we recommended
the Canadian extension (dubbed CanCON) use the QHY camera instead of the
current MallinCAM video camera.  The processing is much simpler than for video
data since the direct product of data collection is to write a series of FITS
images, one per integration.  These cameras have no appreciable dark current
when operated with its cooler set to 0\mydeg C or less.  The readout bias level has a very stable
mean value with structured noise superimposed.  The consequence of this is for
any image that is read-noise limited, you see a low-level horizontal banding
in the image.  This banding has an amplitude of a few counts.  We remove this
signature from the images by subtracting a robust mean from the image on
a row-by-row basis.  When complete, the bias, the readout pattern, and the
sky have all been subtracted leaving a mean background of zero.  The
integration mid-time is then computed from the provided start time and the
integration time.  At this point, the images are ready for the standard
analysis.

\subsection{Lightcurve extraction}

These images were all processed with aperture photometry methods described
in \citet{Buie1992a}.  For each image, an object aperture radius is selected
along with an inner and outer sky annulus radius that optimizes the signal-to-noise ratio (SNR).
The apertures were adjusted somewhat between sites, driven by variations
in image quality either due to seeing, wind shake, or poor focus.  The three
radii, in pixels, are give here for each station: Penticton (6,20,100),
Anarchist Mt. Observatory (4,20,100), Ellensburg (10,20,100), Sisters (3,20,100),
CPSLO (4,20,100), Wildwood (5,15,100), SwRI (4,20,100),
Searchlight (4,20,100), Parker (3,20,100), Blythe (4,20,100),
and Yuma (4,20,100).  The resulting lightcurves are shown in Fig.~\ref{fig-lc}.
In the figure, all of the lightcurves have been adjusted in time to be
relative to the predicted mid-time of the event based on the RECON prediction.
Each point in the graph corresponds to a single image.  The lightcurves are
ordered by cross-track distance so that the most northerly stations
are at the top.  The cross-track uncertainty range covered relative to the RECON
prediction was from $+0.08\sigma$ to $-0.11\sigma$.

\subsection{Special Cases}

The data from a few of the stations required some extra handling.  The
deviations from standard processing are summarized in this sub-section.

{\em Penticton (C-03):}  The team did not turn on the cooler for its
camera, requiring them to take calibration dark images.  These dark
calibration data were used
to remove hot pixels.  The calibration was not ideal 
as the sensor temperature was unregulated during the capture. 
However, this did not appear to
compromise the occultation result in any measured way.

{\em Ellensburg (2-19):} There was an issue with the
images where all sources appeared elongated.  The observers reported
vibration on the observing platform and the data show precisely
the same elongation amount and direction throughout the entire video.
We compensated for this problem by increasing the photometry aperture size.
There was also light haze at this site and with the nearby moon there
was a higher sky background level.  All of these conditions conspired to
elevate the noise level in these data.  Some short dropouts are seen in
this lightcurve but in all cases the target star is still seen in the
image during the dropouts.

{\em Wildwood (V-07):} These data are plagued with issues that could not be fully
resolved.  Looking at the plot in Fig.~\ref{fig-lc}, there is a change in
the noise properties just after the mid-point of the observing sequence.  The team noted
the use of SENSEUP=128, but the data after the noise change appeared to be
consistent with an SENSEUP of 90.  This latter half of the data is consistent in
this regard but the first half of the data have a variable SENSEUP
throughout, starting around 40 and increasing to the value of 90
seen later.  What makes this so mysterious is that the possible values of
SENSEUP do not include 90 and this is controlled by the camera firmware.
There is no way to specify a non-standard setting.  However, we do see
a variable SENSEUP after changing this setting, but usually the camera
has stabilized to the new value in a minute or two.  The constraints
provided by this one dataset are therefore weaker than would otherwise be
the case.  These data would show a central chord if present, but sensitivity
to grazing chords or small secondary bodies was reduced.

{\em Yuma (2-15):} The images were not well-focused at the beginning and also
showed signs of aberration. This resulted in a slight increase in noise
compared to other sites.  The dip at the start of the recording seen in
Fig.~\ref{fig-lc} is due to the observing team installing a dew shield
while collecting data in an attempt to reduce scattered light from the Moon.
The dip observed at 5:23 UTC (+50 seconds relative to prediction) was produced by telescope movements.

\begin{center}
\includegraphics[scale=0.5]{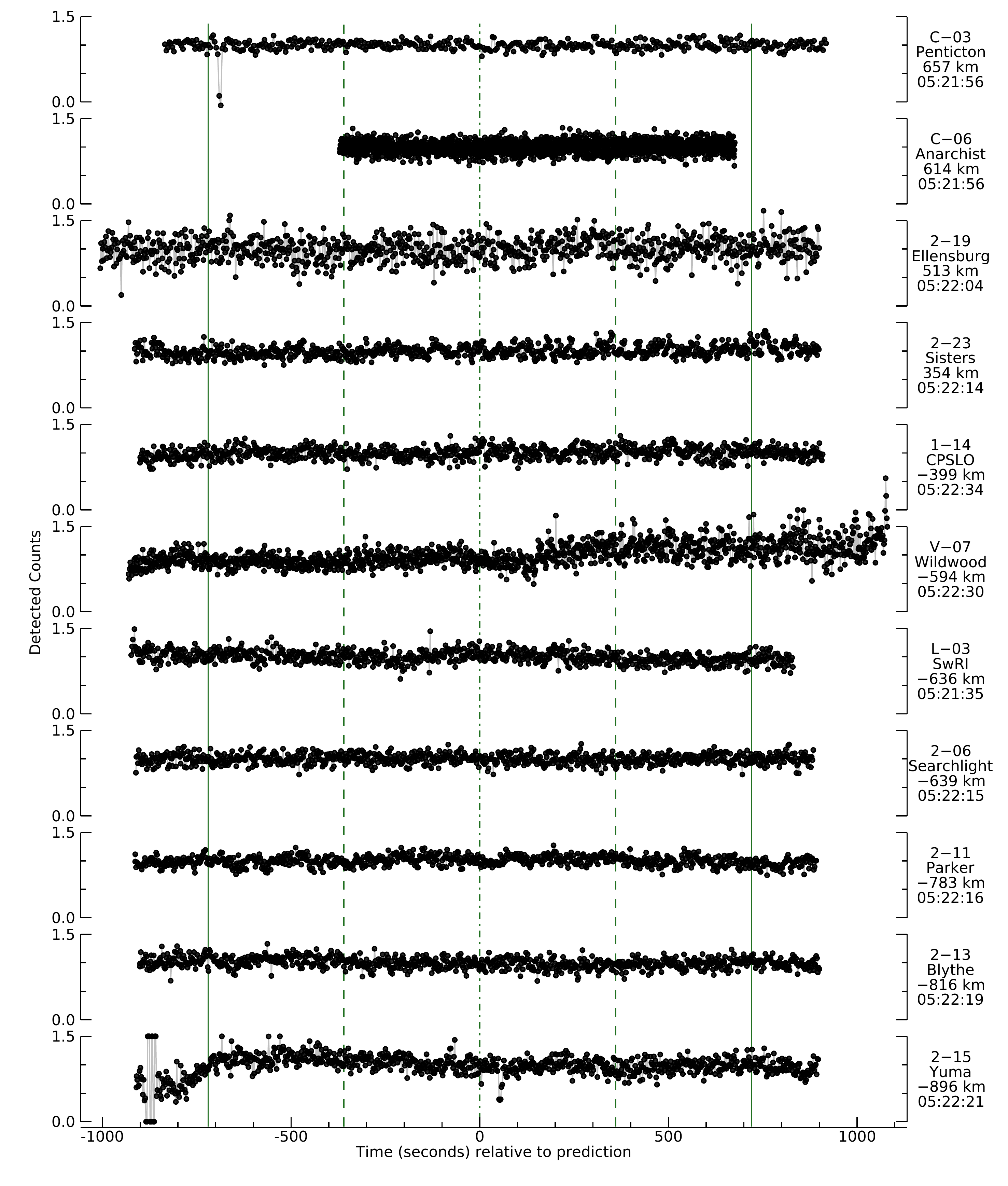}
\figcaption{\label{fig-lc}
Observations from 2018-10-20 occultation.
The figure shows the lightcurves from the data
collected by the RECON stations.  Each sub-plot is labeled
on the right with the team name, cross-track offset, and predicted
event time.
The plots are all normalized to unit flux when the star is visible.
The green vertical lines indicate
the predicted 2-$\sigma$ uncertainty limits for the event.
An electronic copy of the data in this figure is provided.
Uncertainties are included in the electronic data but not
shown on the plot for clarity.
}
\end{center}

\section{Results}\label{sec:results}

Inspection of the data revealed an interesting drop out in the Penticton (C-03)
data very close to $-2\sigma$.  No other temporally correlated dips were
seen in other datasets.  This positive detection is worth a more detailed
examination.  Figure~\ref{fig-event} shows a short segment of the Penticton
data around the time 690 seconds before the nominal RECON prediction.
The model curve shown in red is a simple model of a perfectly sharp occultation
with start and stop times that are consistent with the fluxes from the
lightcurve at the time of the transitions.  Basically, this means if one point
zero and the next is full flux, then the transition must happen at the edge
between the two points.
The timing derived from the model occultation curve as shown in Fig.~\ref{fig-event} is a disappearance at
2018-10-20 05:10:24.069 UTC and a reappearance at 2018-10-20 05:10:31.880
UTC.  The star was occulted for a total of 7.811 seconds, equivalent to
187~km given the sky-plane velocity.  These timing values are provided for
completeness but were not directly used in the diameter determination.

\begin{center}
\includegraphics[scale=0.5]{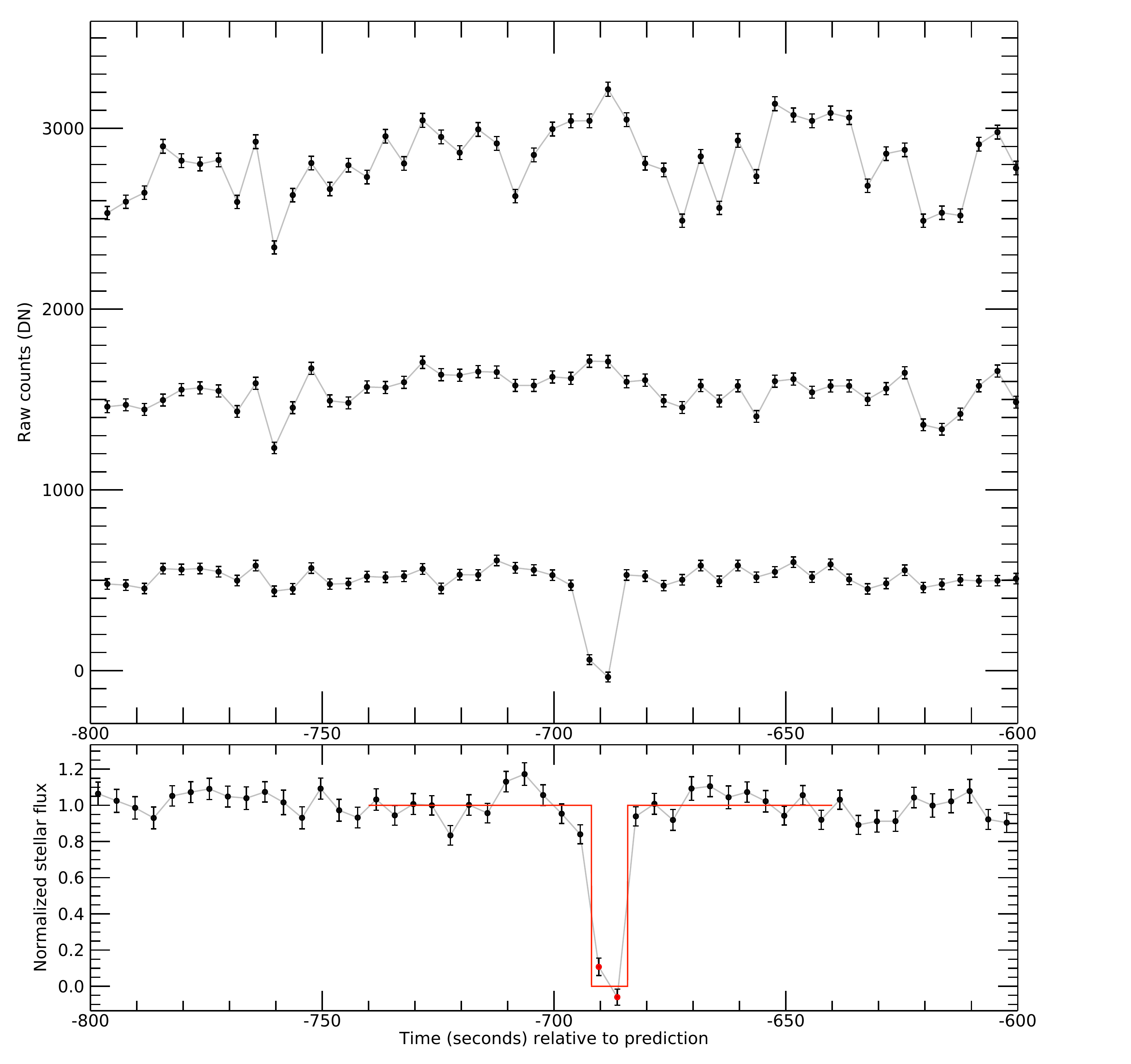}
\figcaption{\label{fig-event}
Detailed view of occultation from Penticton (C-03).
The top panel shows the raw light curves for the target
star (bottom), comparison star 1 (middle), and comparison star
2 (top).  The bottom panel shows the ratio of the target star
to comparison star 2.  This ratio is what is plotted in
Fig.~\ref{fig-lc}.  The red points highlight the occulted or partially
occulted points and the red curve shows a simple model of the occultation.
}
\end{center}

From the Penticton and Ellensburg results, we can place useful constraints
on the size of \tgtes\ even though there was only one measured chord.
A Markov-Chain Monte Carlo (MCMC) scheme is adopted to sample the posterior
probability density function (pdf) of the parameters $x$, $y$ and $r$
constrained by the lightcurves with their point-by-point estimated uncertainties.  $(x,y)$
is the offset with respect to the ephemeris at the time of the occultation
measured in the sky plane at the distance of the object, $r$ is the radius of
the object.  All of the analysis and results here are based on the assumption
of a circular profile and the osculating orbital elements for
\tgtes\ used for the analysis are given in Table~\ref{tbl-elem}.  No diffraction effects
are considered due to the long exposure times and relatively low SNR.  Given the large uncertainty in the prediction, the uncertainty
in $(x,y)$ prior to the occultation is approximately constant in the
relevant zone near Penticton, and it is assumed to have a uniform prior
distribution truncated about 2000~km around the main detection.
For the radius
we consider a power-law with slope $q$=4.5 which is motivated by the
power-law in the differential size distribution for TNOs with radius
$r \gtrsim 50$~km \citep{Bernstein2004,Fuentes2008}.  The distribution is
truncated between $r_{min}$=50~km and $r_{max}$=255~km  given by the physically
motivated limit in the albedo $0.04<p_V<1$.

\begin{deluxetable}{cc}
\tablecaption{Osculating orbital elements\label{tbl-elem}}
\tablewidth{0pt}
\tablehead{
\colhead{Parameter}&
\colhead{Value}
}
\startdata
Epoch    & 2018-10-21 00:00:00 UTC \\
M        &  $ 359.337 \pm 0.032 $ (deg) \\
$\omega$ &  $ 118.109 \pm 0.099 $  (deg)  \\
$\Omega$ &  $ 300.868 \pm 0.005 $  (deg) \\
i        &  $ 11.662 \pm 0.000 $ (deg)  \\
e        &  $   0.937 \pm 0.002 $ \\      
a        &  $ 1018.668 \pm 33.042 $ (au)\\
\enddata
\tablecomments{\scriptsize
Osculating orbital elements for \tgtes\ used in the analysis.
M, $\omega$, $\Omega$, i, a, e are the mean anomaly, argument of perihelion, ascending node, inclination, eccentricity and semi-major axis respectively.
}
\end{deluxetable}

Figures~\ref{fig-geom} and \ref{fig-crosstrck} provide a graphical summary
of the geometry and the astrometric constraints derived from the analysis.  The
lower-left panel in Figure~\ref{fig-geom} shows the (posterior) joint pdf
for the positions $x,y$ in the sky-plane while the panels in the diagonal are the normalized marginal pdf's for each parameter.
The sky-plane is a plane perpendicular to the Earth-star line at the distance of the object with coordinate axes (x,y) in the direction of the east and north and with its origin at the ephemeris position of the center of the object \citep{Elliot1978}.
The nominal solution is taken from the peak in the marginal pdf's.  The solid
contour in the joint distribution is the 39.3\% credible interval while the
vertical dashed lines in the marginal pdf's are the 68\% credible intervals
and represent the formal 1-sigma uncertainties.  The lower-left panel shows the
projected locations of the region constrained by the Penticton (upper) and Ellensburg
(lower) data.  The tracks are computed from the topocentric position of \tgtes\ as
seen from each observing location.  The topocentric scale from Penticton at
the event mid-time was 56,371.9 km/arcsec and the length of the segments
indicate the exposure time of the data.  Superimposed are three illustrative
circular profiles compatible with the data, in red-solid lines the nominal
solution while the green-dashed and blue-dotted circles are the two representative
1-sigma solutions.  Given the single detection and the circular profile assumption,
there is an expected correlation between the x and y parameters in the
cross-track direction.  Also, the asymmetry in the marginal pdf's is clearly
due to the strong negative constraint imposed by the Ellensburg data.
Figure~\ref{fig-crosstrck} shows the positional pdf's projected onto the along-track
$x_v$ and cross-track $y_c$ directions.  The constraint is tighter in the along-track
direction ($x_v$) with $\sim$0.2~mas while the uncertainty in the cross-track direction ($y_c$) is $\sim$2~mas.

\begin{center}
\includegraphics[scale=0.7]{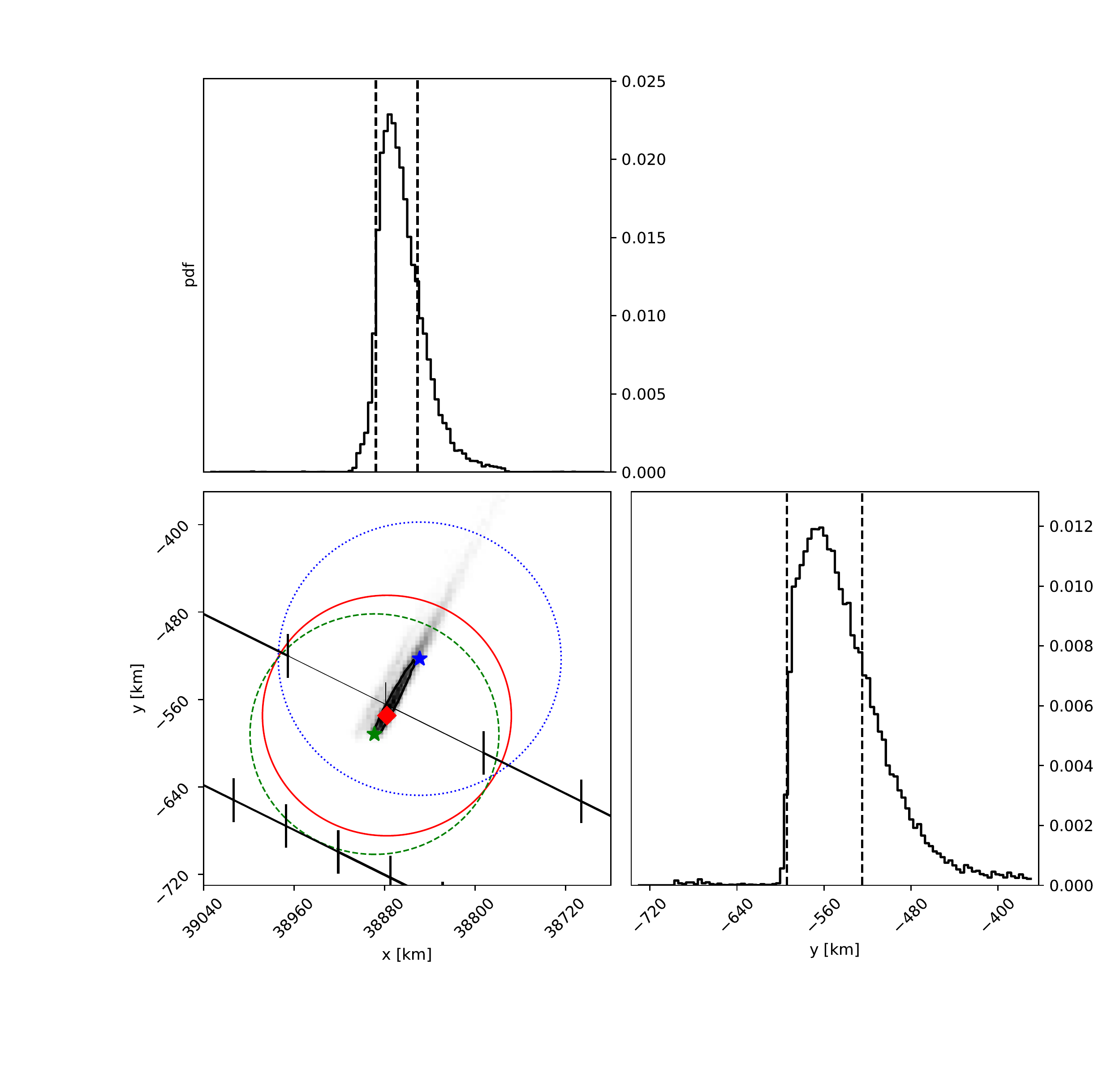}
\figcaption{\label{fig-geom}
Detailed geometry and constraints from the occultation data.  The lower-left
panel is the joint probability density function (pdf) for the offset with
respect to the adopted object ephemeris in the sky-plane.  The contour shows
the 39.3\% credible interval.  The upper and left panels are the normalized marginal
pdf's with the 68\% credible interval defined with dashed vertical lines.
The black segmented lines in the lower-left panel show the tracks for Penticton
(upper) and Ellensburg (lower) with length representing the exposure time and
line thickness indicates the normalized flux.  The occultation detection is shown with
the two central lighter segments.  The solid-red circle (D=220~km) is the
nominal solution for a circular profile centered in the the peak of the
marginal pdf's.  The center is shown with a diamond symbol which is the
adopted astrometric position of the object at reference time $t_{\rm ref}$.
The dashed-green circle (D=220~km) and the large dotted-blue circle (D=250~km)
are the adopted 1-$\sigma$ solutions.  The center indicated by star symbols
are the intersection of the credible interval shown in the marginal pdf's used
to retrieve the astrometric position uncertainties.  The astrometric constraints
are summarized in Table~\ref{tbl-results}.  Notice that the joint pdf extends
to the upper-left whose extreme corresponds to a solution with the imposed limit
$p_V=0.04$ and diameter D=510~km. An electronic copy of the normalized joint pdf in this
figure is provided.
}
\end{center}

\begin{center}
\includegraphics[scale=0.7]{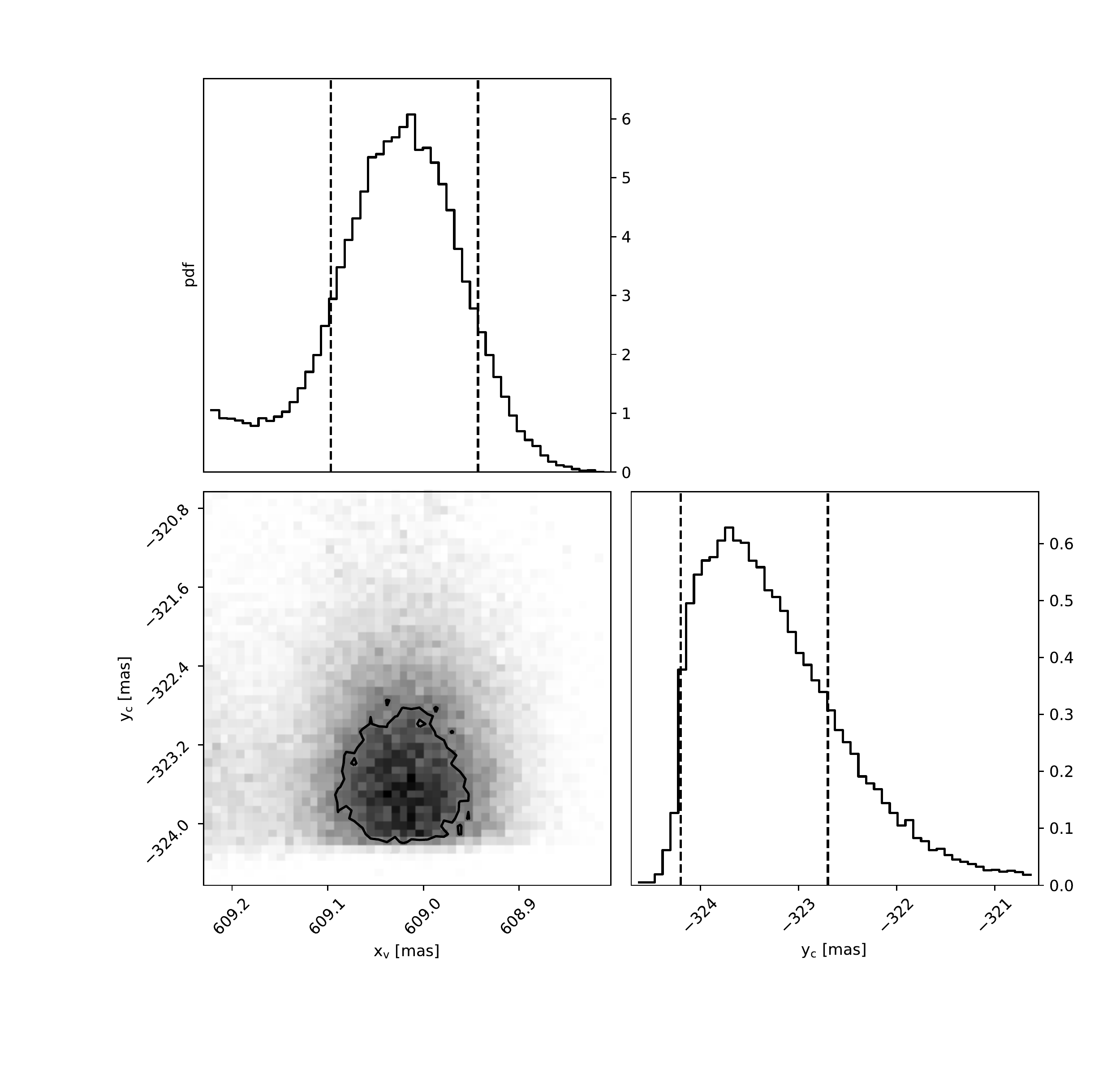}
\figcaption{\label{fig-crosstrck}
Same pdf's from figure \ref{fig-geom} but projected in the along-track $x_v$
and cross-track $y_c$ directions and given in milliarcseconds.  Contours in
the joint pdf's define the 39.3\% credible interval.  The vertical dashed lines
define the 68\% credible intervals.  From here it is observed that the extended
wing in the marginal pdf for the cross-track direction is due to the lack of
constraining occultation data north of the Penticton (C-03) site.  
Notice that, unlike Figure \ref{fig-geom}, the scale of the axes is different and the $x_v$ axis is stretched with respect to the $y_c$ axis.
An electronic copy of the normalized joint pdf in this figure is provided.
}
\end{center}

Each point in the positional pdf implies a size for the object and the
left panel of Fig.~\ref{fig-albedo} shows the pdf for the radius $r$.
Of all the priors in our analysis, the most influential on the
derived size and albedo of the object is the inclusion of a size distribution.
To investigate the sensitivity of our results on the chosen size distribution
we used two additional cases for the priors on the radius.
One extreme case is a prior for $r$ with a uniform distribution between 50 and 255~km.
The other extreme case is a significantly steeper power-law with slope $q$=7
coinciding with one of the more extreme slopes suggested for the Classical KBO population \citep{Fuentes2008}.
The result from the moderate ($q$=4.5) power-law is shown in black and
constrains the object radius to be $r=110^{+14}_{-10}$~km (vertical dashed
lines).  The steeper power-law ($q$=7) shown in red is nearly
identical to that implied by the moderate power-law.  The result from the
uniform prior on size is shown with the dashed-blue lines and exhibits
a long wing in the marginal pdf for larger radii while the location of
the peak in the distribution is essentially the same as the other two cases.
The marginal pdf for the radius $r$ is the main result of our analysis,
subject to the validity of the assumption of a circular profile.

The pdf for the size can then be combined with independent measurements
of the brightness to retrieve a surface albedo.  The right panel of
Fig.~\ref{fig-albedo} shows a case of the albedo implied by an absolute
magnitude of $H_V=5.6$.  For our preferred case of the moderate power-law
the albedo is $p_V=0.21^{+0.03}_{-0.05}$ as indicated by the vertical dashed
lines at the $\pm1\sigma$ locations.  The steep power-law case gives nearly
identical constraints on albedo.  The uniform prior on size shows a bi-modal
pdf distribution in albedo with significant probability given to a low albedo.
We consider the case of a uniform size distribution to be nonphysical.
The inclusion of a size distribution is far more reasonable though the
details of what distribution to use is less important to the determination
of the albedo than obtaining a more accurate measurement of the absolute
magnitude of \tgtes.

\begin{center}
\includegraphics[scale=0.7]{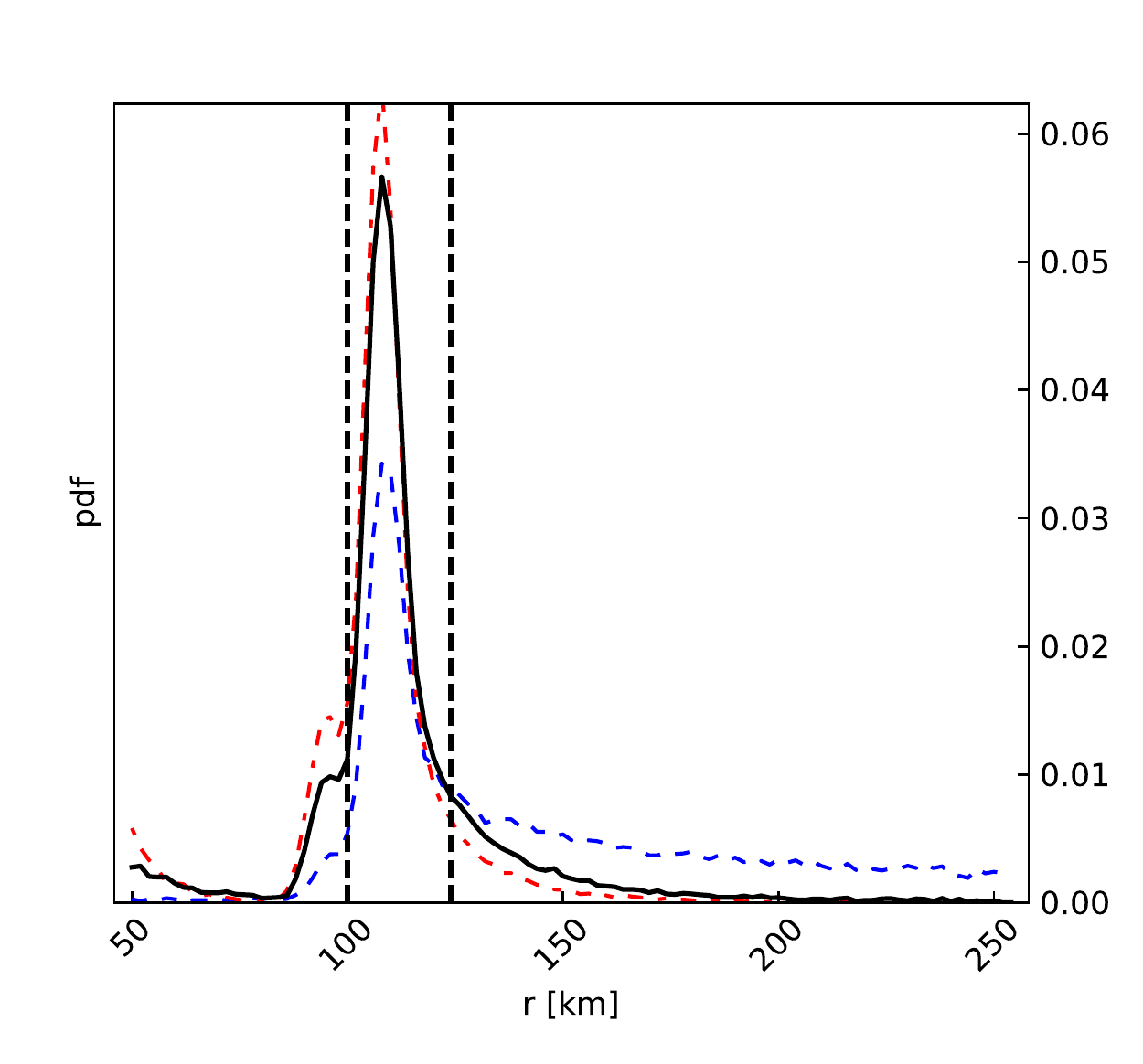}
\includegraphics[scale=0.7]{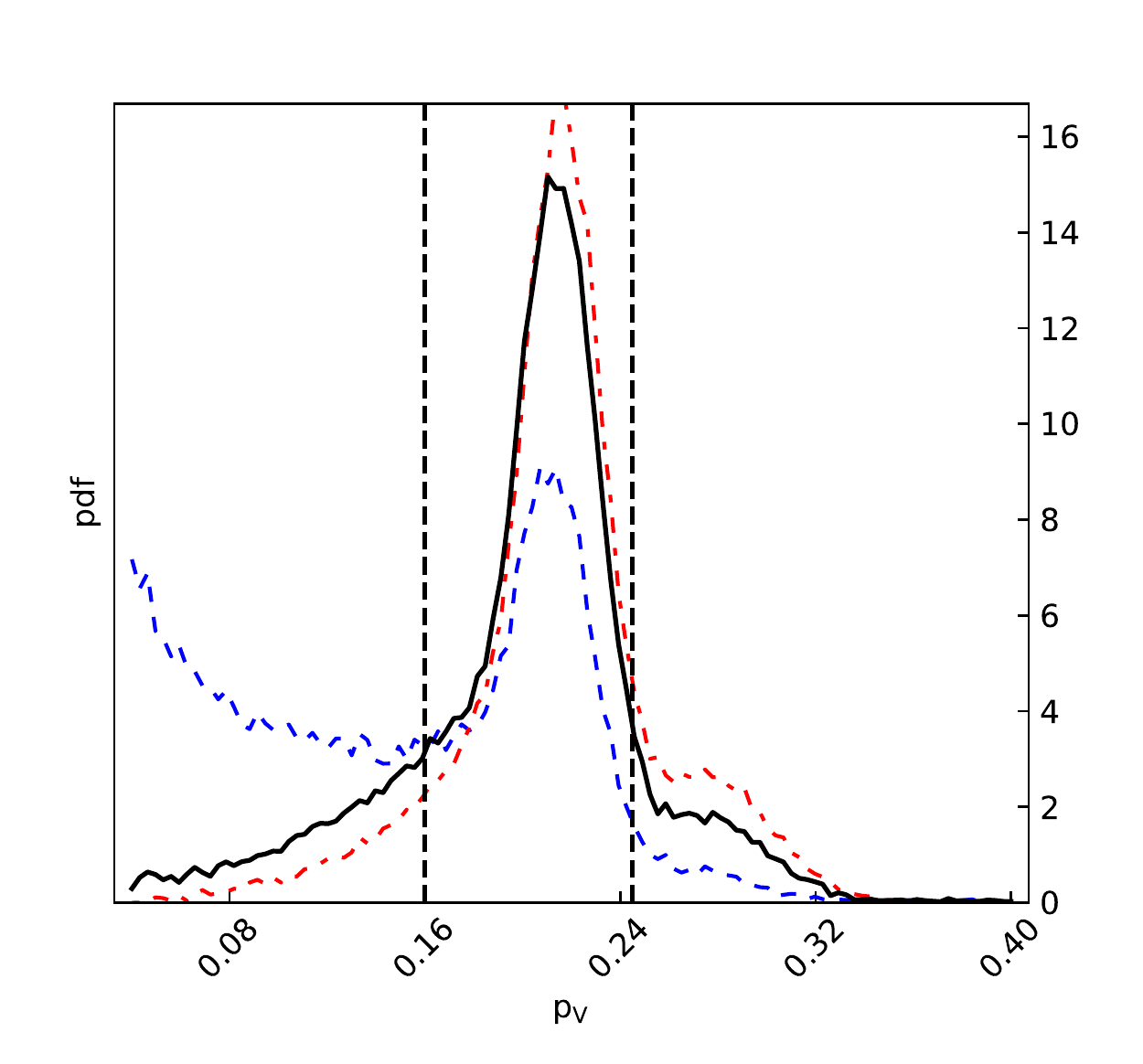}
\figcaption{\label{fig-albedo}
Normalized posterior pdf derived from the occultation analysis comparing different priors for the size distribution.
Left: Posterior pdf's for the object radius $r$.  
Right: Derived posterior pdf for the geometric albedo $p_V$ given the adopted absolute magnitude $H_V$=5.6.
The adopted solution is the solid black curve with a truncated power-law distribution for the prior
in the radius $r$ with a slope $q=4.5$.  
The vertical dashed lines are the 68\% credible intervals for the adopted solution.
For comparison, the dashed-blue lines are the posteriors for a prior in radius $r$ that is uniformly distributed between 50 and 254~km.
The dotted-red lines are the result of using a power-law with a slope $q=7$.
An electronic copy of the marginal pdf for the radius in this figure is provided.
}
\end{center}

Table~\ref{tbl-results} summarizes the astrometric and physical constraints
we derived.  The top section of the table shows the nominal radius $r$
and geometric albedo $p_V$ derived from the occultation analysis adopting an absolute magnitude
$H_V = 5.6$.  The albedo uncertainty does not include any uncertainty
from the absolute magnitude.  A more complete albedo constraint will come
from a better determination of the absolute magnitude of \tgtes\ along with
better uncertainties than can be derived from the MPC database.  A proper
albedo constraint will come from combining our posterior pdf for the size
with the results from more complete photometry.  For now, our results give
a clear indication of the surface having a moderate albedo.

The middle section of Table~\ref{tbl-results} provides the supporting
intermediate values required to compute the final astrometry.  From top to
bottom this begins with the reference time and position of the object.  The
reference time is the time of minimum separation between the star and the object
as seen from the geocenter and using the orbit from Table~\ref{tbl-elem}.  The
ephemeris position ($\alpha_{\rm ref}$,$\delta_{\rm ref}$) at that time is also
given. Next, the measured offset of
the object relative to the ephemeris is given.  This value is the numeric result
shown graphically in Fig.~\ref{fig-geom}.  This offset, when rotated to match
the direction of motion gives the cross-track and down-track offsets ($x_v$,
$y_c$).  The nominal values are from the peak in the marginal pdf's while the
uncertainties are from 68\% credible intervals. Next, using the ephemeris at the reference time and the astrometric position of the star, the offset with respect to the star is also given.  From this final offset, we can than easily compute
the astrometric position $(\alpha,\delta)$ for \tgtes\ at the reference time
$t_{\rm ref}$ with uncertainties given in mas.  These uncertainties do not
include the \gaia\ DR2 positional uncertainty of the star at the epoch of the
event and only include our own measurement errors.  The astrometric position for
\tgtes\ derived in this work will significantly reduce the uncertainties on
future occultation opportunities for this object.

\begin{deluxetable}{cc}
\tablecaption{Astrometric and physical constraints\label{tbl-results}}
\tablewidth{0pt}
\tablehead{
\colhead{Parameter}&
\colhead{Value}
}
\startdata
\multicolumn{2}{c}{Object physical parameters}  \\
r (km)     & $ 110^{+14}_{-10}$ \\
$p_V$($H_V$=5.6)  & $ 0.21^{+0.03}_{-0.05}$ \\
\hline
\multicolumn{2}{c}{Reference time and position of the object} \\
$t_{\rm ref}$ & 2018-10-20 5:22:31.0 UTC \\
$\alpha_{\rm ref}$  &  00:12:17.824253 \\
$\delta_{\rm ref}$  & +13:15:07.43137 \\
\hline
\multicolumn{2}{c}{Offset with respect to the object J2000 ephemeris at $t_{ref}$}  \\
$\Delta \alpha$ $\cos\delta$ (mas) &  689.7$^{+0.2}_{-0.5}$ \\
$\Delta \delta$ (mas) & -10.2$^{+0.9}_{-0.3}$ \\
\hline
\multicolumn{2}{c}{Offset with respect to the object J2000 ephemeris at $t_{ref}$}  \\
\multicolumn{2}{c}{in the along-track and cross-track direction}  \\
$x_v$ (mas)  & $ 609.0 \pm 0.1$  \\
$y_c$ (mas)  & $-323.7 \pm 0.8$ \\
\hline
\multicolumn{2}{c}{Offset with respect to astrometric star position}  \\
$\Delta \alpha$ $\cos\delta$ (mas) & $-294.4^{+0.2}_{-0.5}$ \\
$\Delta \delta$ (mas) & -75.6$^{+0.9}_{-0.3}$ \\
\hline
\multicolumn{2}{c}{Object position at $t_{ref}$ derived from the occultation}  \\
$\alpha$        &  00:12:17.871489 $^{+0.2 mas}_{-0.5 mas}$ \\      
$\delta$        & +13:15:07.42118  $^{+0.9 mas}_{-0.3 mas}$ \\
\enddata
\tablecomments{\scriptsize
Object position and offset with respect to the astrometric position of the star.
$\Delta \alpha \cos(\delta)$,$\Delta \delta$ is given in a J2000 coordinate frame.
Geometric albedo $p_V$ is
based on an adopted absolute magnitude of $H_V=5.6$.
}
\end{deluxetable}

Figure~\ref{fig-dvspv} compares the diameter and albedo of \tgtes\ with
other TNOs.  The figure shows physical data published through 2018 March
from \citet{Johnston2018a} and are highlighted based on dynamical class using
the DES classification system \citep{Elliot2005a}.  The pdf from
Fig.~\ref{fig-albedo} is shown as a black curve for the 1-$\sigma$ region
and a dashed grey curve for the full extension of the pdf.  Based on its
mostly likely albedo from the pdf, \tgtes\ has an albedo comparable to other
Scattered-Extended and Scattered objects.
The average geometric albedo $\overline{p_V}$ for these two dynamical classes 
in the same size range is 0.17 and 0.20 respectively
and 0.21 and 0.22 when all objects, irrespective of diameter, are considered.

\begin{center}
\includegraphics[scale=0.9]{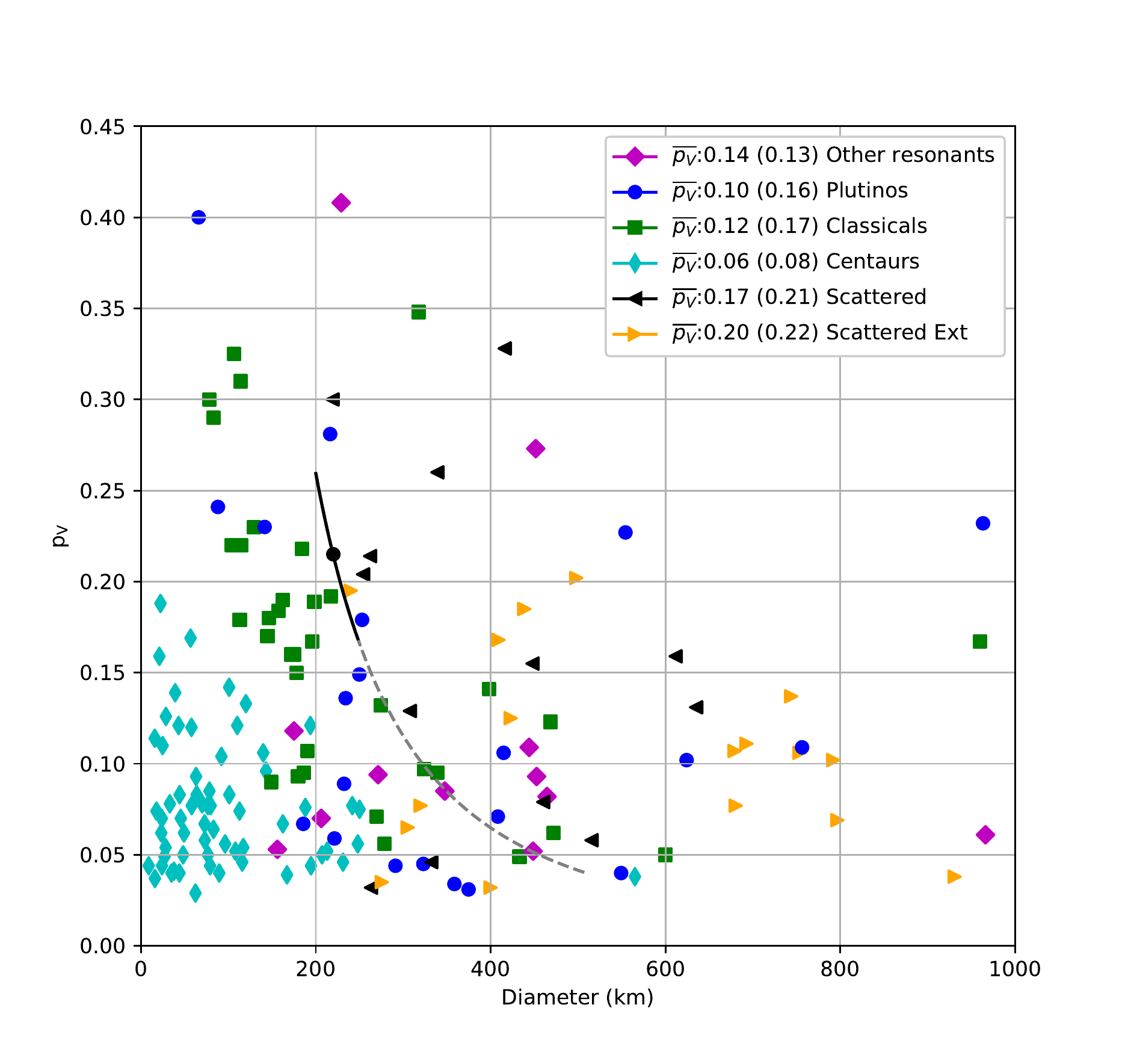}
\figcaption{\label{fig-dvspv}
Size and albedo of \tgtes\ compared to the TNO population.  Diameter and albedo
are taken from a compilation of values through 2018 March from
\citet{Johnston2018a}.   The average value is used for objects with multiple
entries.  The uncertainties are omitted for clarity.  The average geometric
albedo $\overline{p_V}$ for objects in the diameter range $200<D<500$~km is
indicated in the legend.  The average geometric albedo is enclosed in parentheses for all the objects in each class, irrespective of the diameter.
The segmented line is the locus for the adopted absolute magnitude $H_V=$~5.6.  The nominal value and uncertainties
derived in this work are shown with a black dot and solid line respectively.
The albedo for \tgtes\ is comparable to that of Scattered and Scattered-Extended
objects for the same size range.
}
\end{center}

\section{Conclusions}

This first occultation result for \tgtes\ provides a radius between 100~km
and 124~km for a circular profile and suggests an albedo in the range
of 0.16-0.24.  These constraints are the most likely given the
occultation data and a typical size distribution of TNOs.  The albedo
constraint can be improved once better photometry of this object becomes
available.
The geometric
albedo of \tgtes\ is comparable to that of Scattered-Extended and Scattered
objects in the same size range.  Given the assumptions in our analysis and
given our adopted absolute magnitude it appears that we can exclude a low
albedo.  The cumulative probability for the albedo
being $\le5$\% is 0.004.

Of course, the results of this analysis are subject to the validity of our
assumptions.  Given the limited dataset there was little point to
investigating a large number of alternative assumptions.  Nonetheless,
future observations would do well to remember the assumptions we had to make.
Our analysis
shows that the choice of size distribution (other than a uniform prior)
makes little difference.  However, if the object were to have a non-circular
profile at the time of this occultation the change in the inferred albedo
would be significant relative to our derived uncertainties of the circular
case.  If our assumption of a single object is incorrect it could lead to an even bigger
change in the derived results.  For instance, if this object were actually
an equal size and equal albedo binary our data would then apply to just one component and the
projected area of the pair would be twice as big, thus implying an
albedo that is a factor of two lower.  Given our nominal albedo constraint,
such a factor of two reduction would still not imply an unrealistic albedo.
We have not attempted to calculate the likelihood of these
other assumptions but follow up occultation observations might well want to
think about the implications of these unconstrained alternate options when
building a new deployment strategy.  All of these questions can be solved
with additional occultation observations, preferably with far more than just
one positive detection.

These data also provide a new
high-precision astrometric constraint that is a factor of 50 better than
the positional uncertainty at the time of the occultation.  This
uncertainty is comparable to the angular size of the body for the
high-albedo end of the range and will help pave the way for future
occultations with enough chords to get a full measure of its shape.
For future occultation efforts, we recommend a spacing of no more than 90~km
between stations but tighter spacing will be highly desirable.  The spread
of stations would need to be much larger than our measured size to
be sensitive to a binary object.  A second, more
detailed, set of occultation observations can greatly constrain the future
interpretation of this occultation result.

Despite the new astrometric constraint, this object will require further
astrometric observations to preserve the ability to get a high-precision
occultation prediction.  The new measurement provided here does not
appreciably extend the observational arc for this object.  As such, it
merely provides a very accurate fiducial point for the orbit estimate.
The errors in the mean motion will quickly begin to dominate future
predictions without further data.  New astrometry, even at lower precision
than provided by an occultation will continue to improve on the mean motion
just by extending the temporal arc of the data.  However, it would be well
advised for future astrometrists to get high SNR detections.  The best
astrometry being reported from the ground today, reduced against \gaia\ DR2
and measured by the post-orbit fit scatter,
seems to be good to roughly 0.1 arcsec and more data as good as this (or better)
would really help.  Current data on \tgtes\ show an rms scatter of
0.17 arcsec with a third having residuals as large as 0.3-0.4 arcsec
so there is a lot of room for improvement.
We can expect to have future opportunities to learn more about this
distant object provided regular high-quality astrometric
observations are obtained in addition to additional new occultation data.

\begin{acknowledgements}

This work is made possible by the people that are part of RECON and CanCON:
teachers, students, and other community members, including Toochi Brown,
David Bryan Barrows, Ceravolo Anarchist Mt. Observatory, Anna Chase, Dorey Conway,
Mark Cordero, Ron Crawford, Maritz Cruz, Michelle Dean, Anne-Marie Eklund,
Dawn L. Estes, Kristof Klees, Aundriana Magaw, Lara Matheny, Ian R. Norfolk,
Rhonda Palmquist, Nidhi R. Patel, Robert Reaves, Paul Snape, Ihsan A. Turk,
Hunter VandenBosch, Eric Verheyden, Kellen Werts, and Charlene Wiesenborn.
The observers listed in the
table are but a small fraction of the total network and their dedication to this
project is deeply appreciated.  Funding for RECON was provided by a grant
from NSF AST-1413287, AST-1413072, AST-1848621, and AST-1212159.  Some of
the work leading to these results has received funding from the European
Research Council under the European Community's H2020 2014-2020 ERC Grant
Agreement \#669416 ``Lucky Star''.  JJK acknowledges the support of the
Natural Sciences and Engineering Research Council of Canada (NSERC)
[RGPIN/5499-2016].  TJB acknowledges the support of Okanagan College through
the Grants-in-Aid fund.

\end{acknowledgements}

\bibliographystyle{aasjournal}
\bibliography{references}

\end{document}